\documentstyle[prd,aps,preprint,tighten,floats]{revtex}
\begin{document}
\draft
\preprint{IASSNS-HEP-95/112, UPR-687-T, hep-th/9512188}
\date{December 1995}
\title  {Massless BPS-Saturated States on the Two-Torus Moduli Sub-Space 
of Heterotic String}
\author{ Kwan-Leung Chan$^{1}$  and Mirjam Cveti\v c$^{2}$
\thanks{On sabbatic leave from the University of Pennsylvania.} 
}
\address{$^1$\ Department of Physics and Astronomy \\ 
          University of Pennsylvania, Philadelphia PA 19104-6396,\\
           $^2$\  Institute for Advanced Study\\
           Olden lane, Princeton, NJ 08540\\ }\maketitle
\begin{abstract}
{Within a four-dimensional, toroidally compactified heterotic
 string, we identify  (quantized)  charge vectors of electrically charged
BPS-saturated states (along with the  tower of $SL(2,Z)$ 
related dyonic  states), which preserve $1\over 2$ of  $N=4$ supersymmetry and become
massless along  the hyper-surfaces of enhanced  gauge symmetry of the
two-torus moduli sub-space.   
In addition, we identify   charge vectors  of  the  dyonic
BPS-saturated states (along with the  tower of $SL(2,Z)$ related states),
which  preserve ${1\over 4}$ of $N=4$ supersymmetry, and become massless 
at  two points  with the  maximal gauge symmetry enhancement. 
}
\end{abstract}
\newpage

BPS-saturated  states  of four-dimensional, toroidally compactified
heterotic string theory  provide a fruitful ground to  address
non-perturbative aspects of string theory.  In particular, regular solutions,
{\it i.e.}, those  with regular horizons,   may shed light  on  quantum
aspects of black hole physics,
{\it e.g.}, on  statistical interpretation of  black hole entropy \cite{LW,CTII} 
 while those  that can become massless\cite{BEHR,CYII,KL}
at certain points of moduli space may shed light on the nature of enhanced
symmetries \cite{HTII,CYII} at special points of moduli space. Since the effective theory
possessed $N=4$ supersymmetry, the  ADM mass  for these 
BPS-saturated states is protected from quantum corrections. 
In principle, one should be able to trust the BPS mass formula  even in the case
where quantized charges   are of ${\cal O}(1)$.

In this letter  we further address  massless BPS-saturated states. In particular, 
we identify the  (quantized) charge vectors  and  the points (lines,
hyper-surfaces)  in the  moduli space for which  the BPS-saturated
states become massless.
For the sake of simplicity we confine  this study to the  two-torus 
moduli sub-space.  The work generalizes  that of Ref.
 \cite{CYII}, where  the case of  the two-circle 
moduli sub-space was addressed.

Recently,  the explicit form  of the  
generating solution  \cite{CTII,CYIII}  for 
{\it all} the static, spherically symmetric BPS-saturated
states in this class  was obtained.\footnote{In Ref. \cite{CYIII} also all the non-extreme
solutions were obtained. In Ref. \cite{CTII} it was shown that
BPS-saturated  generating solution is an exact   target-space  background  solution 
of a conformal $\sigma$-model. } The generating solution is specified by five
(electric and magnetic) 
charges  of the two  $U(1)_{a,b}$  Kaluza-Klein and two $U(1)_{c,d}$ 
two-form fields  associated with the two, say the first two,
 (toroidally) compactified  dimensions. The most general BPS-saturated state
in this class is  parameterized by 
unconstrained 28 electric
and 28 magnetic charges  
 and is obtained  by  applying a subset of
$T$-duality and $S$-duality transformations, which do not affect the 
four-dimensional space-time, on the generating solution.

The ADM mass  for these states (BPS mass formula), which in general
preserve only  ${1\over 4}$ of supersymmetry, is  specified \cite{CYI,DLR},
in terms  of 28 electric and 28 magnetic charges.
For the purpose of studying the  moduli  (and
the dilaton-axion)  dependence of the BPS mass  
formula\cite{CYI,DLR} we rewrite  it  in  terms of {\it conserved}
 magnetic
($\vec\beta$)  and electric ($\vec\alpha$) 
charge vectors \cite{CTII}: \footnote{
We use the notation  and conventions,  as  specified  in  Refs. \cite{CYIII},
following {\it e.g.},  Ref. \cite{MS,SEN}.}
\begin{equation}
M_{BPS}^2 ={\textstyle{1\over 2} }e^{-2\phi_\infty} 
{\vec \beta}^T \mu_R{\vec \beta} +
{\textstyle{1\over 2}}e^{2\phi_\infty}{\vec {\tilde\alpha}^T}
 \mu_R{\vec {\tilde\alpha}}
+ \left [({\vec \beta}^T\mu_R{\vec \beta})
({\vec \alpha}^T\mu_R{\vec \alpha})-({\vec \beta}^T\mu_R
{\vec \alpha})^2\right]^{1\over 2},
\label{ME}
\end{equation}
where 
\begin{equation}
{\vec {\tilde \alpha}}\equiv{\vec \alpha}+\Psi_{\infty}{\vec\beta}, \ \ 
\mu_{R,L}\equiv M_{\infty}\pm L.
\end{equation} 
The charge vectors $\vec \alpha$ and $\vec \beta$   are related to the
physical magnetic  $\vec P$ and electric $\vec Q$ charges  in the
following way:
\begin{equation}
\sqrt{2}P_i=L_{ij}{\beta}_j\ , \ \ \ \ \sqrt{2}Q_i = e^{2\phi_{\infty}}
M_{ij\,\infty}(\alpha_j + \Psi_{\infty}\beta_j),\ \  (i=1,\cdots , 28)
\label{charges}
\end{equation}
where the subscript
 $\infty$ refers to the asymptotic ($r\to\infty$)
  value of the  corresponding fields.  Here, the  
 moduli matrix $M$ 
and the dilaton-axion field
$S\equiv\Psi+i{\rm e}^{-2\phi}$, transform covariantly
(along with the corresponding charge vectors) under the $T$- duality 
($O(6,22,Z)$) and
$S$-duality ($SL(2,Z)_S$), respectively,  while the BPS mass formula (\ref{ME})   remains
invariant under these transformations.

Note that when the magnetic  and electric  charges
are parallel in the $O(6,22)$ sense, {\it i.e.},  $\vec{\beta}\propto \vec{\alpha}$
(in the quantized theory the lattice charge  vectors should be relative
co-primes \cite{SEN}),
the BPS mass formula (\ref{ME})  is that of the
BPS-saturated states which  preserve ${1\over 2}$ of $N=4$ supersymmetry
(see, {\it e.g.}, \cite{SEN}).   In the case when the magnetic and
electric charges are not parallel, the mass is larger (the last term in
(\ref{ME}) is non-zero) and the
configurations preserve only $1\over 4$ of $N=4$ supersymmetry \cite{CYI}.
Note that states  preserving ${1\over 2}$ of $N=4$ supersymmetry belong to
the vector super-multiplets, while those  preserving ${1\over 4}$ of $N=4$
supersymmetry belong to the highest spin ${3\over
2}$-supermultiplets \cite{STRAD,KALL}. Thus, when the former [latter] states become
massless they may contribute to the enhancement of gauge 
symmetry\cite{HTII} [supersymmetry\cite{CYII}].

In the quantum theory  the charge vectors $\vec{\alpha}$, ${\vec\beta}$  are 
quantized. Following \cite{SEN},
one may attempt to constrain the allowed lattice charge vectors  by using
the constraints  for the elementary
 BPS-saturated string states of   toroidally compactified heterotic string,
 along with the  Dirac-Schwinger-Zwanziger-Witten (DSZW) \cite{WITTENIII}
 quantization condition.\footnote{In Ref. \cite{CTII} 
the charge quantization for the generating solution
is implied  by considering the conformal field theory  describing the throat region  of the
corresponding string solution.}
Purely electric BPS-saturated states (${\vec\beta}=0$)
 preserve $1\over 2$ of $N=4$ supersymmetry
and  have the same quantum numbers \cite{DR,CMP}
as the  elementary   BPS-saturated string states with
no excitations in the right-moving sector
 ($N_R={1\over 2}$). For the  electric   states  the 
quantized    charge vector $\vec\alpha$  is then  constrained to lie on an even
self-dual   lattice  $\Lambda_{6,22}$ 
with  the  following   norm (in the $O(6,22)$ sense)
\cite{SEN}:
\begin{equation}
{\vec \alpha}^TL{\vec\alpha}=2N_L-2=-2,0, 2, ... \ ,
\label{chl}
\end{equation}
where  the integer $N_L$  parameterizes the  level of the left-moving sector.
 
The  DSZW  charge quantization condition then  implies
an analogous  constraint for
${\vec \beta}^TL{\vec\beta}$; magnetic
 charge vectors $\vec \beta$ are  then  constrained  \cite{SEN} 
to lie on an even self-dual
   lattice  $\Lambda_{6,22}$  with the norm
\begin{equation}
{\vec \beta}^TL{\vec\beta}=2N_L-2=-2,0, 2, ... \ .
\label{chlII}
\end{equation}
Since we confine the analysis to 
 the two-torus moduli  sub-space , the $T$-duality group reduces to 
$O(2,2)$. Then only  the $O(2,2)$  part of the   symmetric moduli metric $M$
is non-trivial and of the form:
\begin{equation}
M=\left ( \matrix{G^{-1} & -G^{-1}B \cr 
-B^T G^{-1} & G + B^T G^{-1}B} 
\right ), \ \ \  L =\left ( \matrix{0 & I_2\cr
I_2 & 0} \right )
\label{modulthree}
\end{equation} 
where $G \equiv [{G}_{mn}]$ ($(m,n)=1,2$), $B\equiv B_{12}$ 
 are the four moduli of
the two-torus and $L$ is an $O(2,2)$ invariant matrix.

The four moduli fields can be expressed in terms
of two complex fields $T$ and $U$, (see, {\it e.g.}, 
 \cite{GPR} and references therein):
\begin{equation}
{T \equiv \sqrt{{\cal G}} + i B}, \ \ \ 
U \equiv {{\sqrt{\cal G} - i G_{12} }\over G_{11}}
\end{equation}
where ${\cal G} \equiv {\rm det}(G_{mn})$.  The $T$ and $U$ fields transform 
 covariantly under  $PSL(2,Z)_T $ and ${ PSL(2,Z)}_U$,
respectively, {\it i.e.}, the subgroups of the
duality  group  $O(2,2,Z)={ PSL(2,Z)}_T \times { PSL(2,Z)}_U
\times Z_{2\, T\leftrightarrow U} $ \cite{GPR}.

In order to address massless BPS states which preserve $1 \over 2$ of
supersymmetry, we first concentrate on purely electrically charged configurations
($\vec\beta=0$) with  the  electric lattice  charge vector 
${\vec \alpha}\equiv (\alpha_a,\alpha_b;\alpha_c,\alpha_d)$ 
 whose norm  is:  
\begin{equation}
{\vec\alpha}^TL{\vec\alpha}=-2.
\label{elc}
\end{equation}
Namely, only the states  with  the  electric charge 
norm  (\ref{elc})  can become massless\cite{HTII} along the lines
(hyper-surfaces) of moduli space
for which:
\begin{equation}
{\vec\alpha}^T\mu_R{\vec\alpha}=0.
\label{zmc1}\end{equation}
It turns out that  (\ref{zmc1}) is satisfied along the following hyper-surfaces,
along with the following 
 accompanying electric charge vectors ${\vec \alpha}$:\footnote{In the following we
 suppress the subscript $\infty$  for the asymptotic values of the  moduli
 fields.}  
\begin{equation}
{{\cal L}_1: U = T\Leftrightarrow(G_{11},G_{22},G_{12},B)=(1,G_{22},-B,B);
\ \ {\vec\alpha}= {\vec {\lambda}}_{1\,\pm}\equiv\pm (1,0,-1,0),}
\label{line1}
\end{equation}

\begin{equation}
{{\cal L}_2: U = {1 \over T} \Leftrightarrow 
(G_{11},G_{22},G_{12},B)=(G_{11},1,B,B)};\ \ 
{\vec\alpha}={\vec {\lambda}}_{2\,\pm}\equiv\pm (0,1,0,-1),
\label{line2}
\end{equation}

\begin{equation}
{{\cal L}_3: U = T - i \Leftrightarrow
(G_{11},G_{22},G_{12},B)=(1,G_{22},1-B,B)}; \ \ 
{\vec {\alpha}}={\vec{\lambda}}_{3\,\pm}\equiv\pm (1,1,-1,0),
\label{line3}
\end{equation}

$$
{{\cal L}_4: U = {T \over {i T + 1} } \Leftrightarrow
(G_{11},G_{22},G_{12},B)=(G_{11},-1+2B+G_{11},-1+B+G_{11},B);}$$
\begin{equation}
{\vec {\alpha}}={\vec{\lambda}}_{4\,\pm}\equiv\pm (1,0,-1,1),
\label{line4}
\end{equation}
Those  are the {\it same}  four hype-surfaces of the two-torus moduli sub-space  
(in the fundamental domain),  for which
the gauge symmetry
of toroidally compactified  heterotic string is 
enhanced due to the Halpern-Frenkel-Ka\v c mechanism, {\it i.e.},  those are
the hyper-surfaces where the perturbative string states (with
$N_R={1\over 2}$), which have the same quantum numbers as electrically charged
BPS-saturated states, become massless.
Thus, on the heterotic side these electrically charged 
BPS-states  are identified with the
elementary string excitations. 

Along  each of the  hyper-surfaces ${\cal L}_{1,2,3,4}$ these  electrically
charged massless BPS-saturated states,  which are
 scalar components of the  vector super-multiplets,
contribute to the enhancement of the  gauge symmetry from 
$[U(1)_a\times U(1)_b\times U(1)_c\times U(1)_d]$  (at generic points of moduli
space)  to 
 $[U(1)_b\times U(1)_d \times U(1)_{a+c}\times
SU(2)_{a-c}]$, $[U(1)_a\times U(1)_c \times U(1)_{b+d}\times SU(2)_{b-d}]$,
$[U(1)_d\times U(1)_{a+c} \times U(1)_{a-2b-c}\times SU(2)_{a+b-c}]$ and 
$[U(1)_b\times U(1)_{a+c}\times U(1)_{a-c-2d}\times SU(2)_{a-c+d}]$, respectively.

 At  the point $U=T=1$, {\it i.e.}, $(G_{11},G_{22},G_{12},B)=(1,1,0,0)$ (the self-dual point of
the two-circle),   ${\cal L}_1$  and  ${\cal L}_2$  meet and 
the enhanced  gauge symmetry is 
$[U(1)_{a+c}\times U(1)_{b+d}\times SU(2)_{a-c}\times
SU(2)_{b-d}]$. At  the point $T=U^*=e^{i {\pi \over 6}}$,
{\it i.e.}, $(G_{11},G_{22},G_{12},B)=(1,1,{1 \over 2},{1 \over 2})$, 
 ${\cal L}_2$, ${\cal  L}_3$  and ${\cal L}_4$ 
 meet and the enhanced  gauge symmetry  is $[U(1)_{a+c}\times
U(1)_{a-2b-c-2d}\times SU(3)_{b-d,2a+b-2c+d}]$.  Here
the subscript(s) for the
non-Abelian gauge factors  ($SU(2)$, $SU(3)$) denote the linear 
combinations  of the  Abelian  generators  that determine
the diagonal  generator(s) of the non-Abelian factors.

Due to the $SL(2,Z)_S$ symmetry, along with
each of the charge vectors  ${\vec \alpha}$ (as specified in
(\ref{line1})-(\ref{line4})), there is a  tower of dyonic configurations  
(including the $Z_2$  related  purely magnetic states)
with $p{\vec \beta}=q{\vec\alpha}$, where  $p$ and $q$ are the relative
co-primes \cite{SEN}. 
 These dyonic configurations 
 become massless at the same points of moduli space as
purely electric configurations. 

We now address massless dyonic states  
whose electric and magnetic charge vectors {\it are not parallel}. 
These states only preserve $1\over 4$ of $N=4$ supersymmetry.  The necessary
condition for them to become massless is that
 both the  electric and the  magnetic  charge vector
 norms satisfy:
 \begin{equation}
{\vec\alpha}^TL{\vec\alpha}=-2, \ \  
 {\vec\beta}^TL{\vec\beta}=-2.
\label{lcv}\end{equation}
These BPS-saturated states become massless
at  the points of moduli space  for which now  the following {\it three} constraints are satisfied: 
\begin{equation}  
{\vec\alpha}^T\mu_R{\vec\alpha}=0,\ \ {\vec\beta}^T\mu_R{\vec\beta}=0,\ 
\ {\vec\beta}^T\mu_R{\vec\alpha}=0, 
\label{const}
\end{equation} 
By explicit calculation we found that  the three constraints (\ref{const})
 are satisfied only at  the  following  two points:
\begin{equation}
T=U=1,  \ \ \ \ \ \ \ \ \ \ \ \ ({\vec \alpha},{\vec \beta})=({\vec\lambda}_{1\,\pm}, 
{\vec\lambda}_{2\,\pm}), \ \ 
\label{dy1}\end{equation}
\begin{equation}
T=U^*=e^{i {\pi \over 6}}, \ \ \ \ \ \ \ \ 
({\vec \alpha},{\vec \beta})=({\vec\lambda}_{i\,\pm},{\vec\lambda}_{j\,\pm}), 
\ \  [(i,j)=2,3,4,\ \ i< j].
\label{dy2}
\end{equation}
The charge assignments  for the four  massless dyonic  BPS-saturated  states
 (\ref{dy1})
at the self-dual point of the  two-circle were found in Ref. \cite{CYII}. 
At the point $T=U^*=e^{i {\pi \over 6}}$ there are
twelve massless dyonic BPS-saturated  states (\ref{dy2}).  In addition, there
is an  infinite $SL(2,Z)_S$ related tower of  massless states (including the
$Z_2$ related states with  electric and magnetic charge vectors  in
(\ref{dy1}) and (\ref{dy2}) interchanged).  Since these states
 belong to the highest
spin $3\over 2$-supermultiplet, they may contribute to the enhancement of
supersymmetry there.
Note that   dyonic states  (\ref{dy1}) and (\ref{dy2}) {\it are not} in the perturbative spectrum of
toroidally compactified heterotic  string.

A few comments are in order. The  discussed BPS-saturated states become massless at 
special points  and hyper-surfaces of moduli space, regardless of the
strength of the dilaton-axion coupling.\footnote{Note, that the BPS mass formula (\ref{ME}) is semi-positive definite for
any asymptotic value of the axion field $\Psi_\infty$. This result is due to
the fact that the  lattice vectors satisfy 
${\vec\alpha}^T\mu_R{\vec\alpha}\ge 0,\ \ {\vec\beta}^T\mu_R{\vec\beta}\ge0$,\ 
and $({\vec\alpha}^T\mu_R{\vec\alpha})( {\vec\beta}^T\mu_R{\vec\beta}) -
({\vec\beta}^T\mu_R{\vec\alpha})^2\ge 0$ everywhere in the moduli space.} 
Note also that all the discussed  states are singular four-dimensional
 solutions. Namely,
for the solutions to be regular, {\it i.e.},  with the (Einstein frame) 
 event  horizon,
the norms of the lattice charge vectors  have to  satisfy  
the following constraints \cite{CTII}:
\begin{equation}
{\vec \alpha }^TL{\vec\alpha }> 0  \ ,\  \ \ \
{\vec \beta}^TL{\vec\beta}> 0\ , \  \ \  \
({\vec \beta}^T L {\vec \beta})
({\vec {\alpha}}^T L{\vec {\alpha}})-({\vec \beta}^T L  {\vec
{\alpha}})^2 > 0.
\label{norm}
\end{equation}
Since the norms  (\ref{elc}), (\ref{lcv}) of the massless BPS states 
 are negative, all the above
solutions  are singular  from the four-dimensional point of view.

The above solutions were obtained as
semi-classical solutions of  toroidally compactified
 heterotic string; they are parameterized in
terms of  classical bosonic fields  of heterotic string 
and (quantized) lattice  charge 
vectors, consistent with the  heterotic string constraints and  the 
DSZW quantization condition.  It is important to address the stability of these
configurations, as well as 
to identify these semi-classical  solutions
 in terms of the $D-$brane\cite{POL}  solutions 
 of   Type IIA  string.

\vskip 0.5in
After the results  presented in the paper had been obtained, 
 we became aware of the paper \cite{CLR}  where  related issues  were addressed.

\acknowledgments
The work was supported in part by U.S. Department of Energy Grant No. 
DOE-EY-76-02-3071, the Institute for Advanced Study funds  and J. Seward
Johnson foundation (M.C.), and the National Science  Foundation Career
Advancement Award  PHY95-12732 (M.C.). M.C. would like to thank D. Youm for
discussions.


\begin{references}

\bibitem{LW}{A. Sen, Mod. Phys. Lett. {\bf A10} (1995), 2081, hep-th/9504147; 
A. Peet, Nucl. Phys. {\bf B456} (1995) 732, hep-th/9506200; 
 F. Larsen  and F. Wilczek, PUPT-1576,  hep-th/9511064.}
\bibitem{CTII}{M. Cveti\v c and  A.A.  Tseytlin,
IASSNS-HEP-95-102, hep-th/9512031.}

\bibitem{BEHR}{K. Behrndt,  Nucl. Phys.{\bf B455} (1995) 188, 
hep-th/9506106 R. Kallosh,  Phys. Rev. {\bf D52}  (1995) 6020, 
hep-th/9506113;  R. Kallosh and A. Linde, Phys. Rev. {\bf D52} (1995) 7137, 
hep-th/9507022.}

\bibitem{CYII}{M. Cveti\v c and D. Youm,  Phys. Lett. {\bf B359} (1995) 87,
hep-th/9507160.}

\bibitem{KL}{R. Kallosh and  A. Linde, SU-ITP-95-26,
 hep-th/9511115.} 

\bibitem{HTII}{C.M. Hull and P.K. Townsend, 
Nucl. Phys. {\bf B451} (1995) 525,  hep-th/9505073.}

\bibitem{CYIII}{M. Cveti\v c and D. Youm, IASSNS-HEP-95/107, hep-th/9512127.}

\bibitem{CYI}{M. Cveti\v c and D. Youm,
 Phys. Rev. {\bf D53} (1996) R584, hep-th/9507090.}

\bibitem{DLR}{M.J. Duff, J.T. Liu and J. Rahmfeld, CTP-TAMU-27-95,
 hep-th/9508094.}

\bibitem{MS}{J. Maharana and J.H.   Schwarz, Nucl. Phys. {\bf B390} (1993) 3,
 hep-th/9207016. }
\bibitem{SEN}{A. Sen, Int. J. Mod. Phys. {\bf A9} (1994) 3707,
hep-th/9402002.}

\bibitem{STRAD}{J. Strathdee, Int. J. Mod. Phys. {\bf A2} (1987) 273.}
\bibitem{KALL}{R. Kallosh, Phys. Rev. {\bf D52} (1995) 1234, hep-th/9503029.}

\bibitem{WITTENIII}{E. Witten, Phys. Lett. {\bf B86} (1979) 283.}
\bibitem{DR}{M.J. Duff and J. Rahmfeld, Phys. Lett. {\bf B345} 
(1995) 441, hep-th/9406105.}
\bibitem{CMP}{ C.G. Callan, J.M.  Maldacena  and A.W. Peet,
PUPT-1565,  hep-th/9510134;
 A. Dabholkar, J.P. Gauntlett, J.A. Harvey and D. Waldram,
 CALT-68-2028, hep-th/9511053. }
\bibitem{GPR}{A. Giveon, M. Porrati and E. Rabinovici, Phys. Rep. {\bf 244}
(1994) 77, hep-th/9401139.}

\bibitem{POL}{J. Polchinski, Phys. Rev. Lett. {\bf 75} (1995) 4724-4727, hep-th/9510017.} 
\bibitem{CLR}{G. Lopes Cardoso, G. Curio, D. L\" ust, T. Mohaupt
 and S-J. Rey, HUB-EP-95-33, hep-th/9512129.}

\end{references}
\end{document}